%% file: Main-file.tex
\documentclass[sigconf,bookmarksnumbered,unicode,colorlinks=true,allcolors=blue]{aamas} 

\usepackage{balance} 

\usepackage{algorithm}
\usepackage{algpseudocode}
\usepackage{xcolor}
\usepackage{balance}

\usepackage{dblfloatfix}

\graphicspath{{figures/}}
\newcommand{\mainfig}[2]{\includegraphics[width=#1]{#2}}

\newcommand{\prop}{proponent}
\newcommand{\chal}{challenger}
\newcommand{\Prop}{Proponent}
\newcommand{\Chal}{Challenger}

\makeatletter
\gdef\@copyrightpermission{
  \begin{minipage}{0.2\columnwidth}
   \href{https://creativecommons.org/licenses/by/4.0/}{\includegraphics[width=0.90\textwidth]{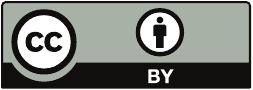}}
  \end{minipage}\hfill
  \begin{minipage}{0.8\columnwidth}
   \href{https://creativecommons.org/licenses/by/4.0/}{This work is licensed under a Creative Commons Attribution International 4.0 License.}
  \end{minipage}
  \vspace{5pt}
}
\makeatother

\submissionType{Research Paper Track}

\title{
Competing Influence in LLM Populations: Budget Timing and Baseline Drift}

\author{Mehwish Nasim}
\affiliation{
  \institution{The University of Western Australia}
  \city{Perth}
  \country{Australia}}
\email{mehwish.nasim@uwa.edu.au}

\author{Nicolas Fay}
\affiliation{
  \institution{The University of Western Australia}
  \city{Perth}
  \country{Australia}}
\email{nicolas.fay@uwa.edu.au}

\begin{abstract}
When competing campaigns target a population of large language model (LLM) agents, observed opinion changes may reflect both campaign influence and changes that would occur without intervention. We study how the timing of campaign expenditure and the population's own evolution shape outcomes in a network of persona-conditioned LLM agents that exchange and revise opinions in natural language. A proponent and a challenger are tasked with generating opposing messages under equal delivery budgets, while a separate stance probe measures opinions without influencing the simulation. Across eight claims, we compare single-campaign and competitive conditions against matched no-campaign runs, and evaluate all nine combinations of early, sustained, and late spending on a reference topic. Baseline stance drift varies substantially across claims and, on some, accounts for a sizeable fraction of apparent campaign effects. After accounting for this drift, the direction of observed campaign-associated changes aligns more closely with the claim's epistemic standing than with the campaigner's assigned role. Uninfluenced populations also consistently lose stance diversity, demonstrating that directional opinion change and convergence are distinct phenomena. In networks with two opinion-segregated communities, campaigns primarily reduce differences between communities rather than diversity within them. LLM campaign agents do not always argue for their assigned positions, particularly when these conflict with established evidence. Finally, campaign outcomes depend on the opponent's expenditure schedule: unlike a matched numerical bounded-confidence population receiving interventions at the same times and locations, the LLM population exhibits a payoff structure with no pure-strategy equilibrium. These findings highlight the importance of matched baselines, campaign-message fidelity checks, and population-level diversity measures when evaluating strategic influence in artificial LLM societies.
\end{abstract}


\newcommand{\BibTeX}{\rm B\kern-.05em{\sc i\kern-.025em b}\kern-.08em\TeX}

\begin{document}




\maketitle 


\section{Introduction}
\label{sec:introduction}

Online opinion formation is shaped by repeated interaction between individuals, their social environments, and external actors seeking to influence beliefs. This is particularly important in misinformation settings, where adversarial actors attempt to shift opinions while corrective interventions seek to limit or reverse their effects. Classical opinion-dynamics models provide a principled way to study such processes by representing individuals as nodes in a network whose beliefs evolve according to predefined interaction rules. Models such as DeGroot~\cite{degroot1974reaching}, Friedkin--Johnsen~\cite{friedkin1990social}, and bounded-confidence models~\cite{deffuant2000mixing,hegselmann2002opinion} have produced important insights into consensus, polarisation, and fragmentation. Their abstraction, however, comes at a cost: opinions are typically represented as scalar values, and the semantic content of the messages through which influence occurs is largely absent.

Large language models (LLMs) provide a different substrate for agent-based social simulation. LLM agents can receive and generate natural-language arguments, maintain persona-dependent behaviour, and revise expressed beliefs in response to contextualised interaction. Recent work shows that such populations can exhibit collective opinion dynamics, including convergence, fragmentation, persuasion asymmetries, and polarisation~\cite{chuang2024simulating,cau2025selective,wang2025decoding}. These dynamics are not behaviourally neutral, however. Networked LLM agents can exhibit model-dependent opinion attractors and systematic topic, agreement, and anchoring biases~\cite{probine2026characterizing,brockers2026disentangling}. Consequently, an observed shift towards one side of a debate may reflect strategic influence, properties of the persuasive messages, or an intrinsic tendency of the underlying model.

This distinction becomes especially important in misinformation settings. Recent LLM-based social simulations show that persona-conditioned populations can be manipulated by adversarial actors and that interventions such as fact-checking, corrective agents, or resistance mechanisms can reduce misinformation or polarisation~\cite{dehkordi2026polarization,liu2025mosaic,borah2026persuasion}. Yet these studies generally treat mitigation as a mechanism or intervention rather than as a competing strategic actor subject to the same constraints as the adversary.

A parallel literature on competitive influence games explicitly studies such strategic interaction. In these models, opposing actors allocate limited resources across individuals or campaign periods while population opinions evolve according to classical numerical rules~\cite{dhamal2020twophase,etesami2022openloop,bastopcu2025online}. This literature shows that influence depends not only on total resources but also on when and where they are deployed. However, persuasive content and context-dependent recipient responses remain outside the model.

Qasmi et al.~\cite{qasmi2025competing} model misinformation and debunking
as a non-cooperative game between resource-constrained LLM agents. Their
competing messages influence a population governed by bounded-confidence
dynamics. Aggressive early expenditure by the debunking agent can produce
short-term gains but leave insufficient resources to sustain them. Whether
this temporal effect persists when the population itself revises beliefs
through natural-language interaction remains unresolved.

In this work, we replace the scalar bounded-confidence population with a network of persona-conditioned LLM agents whose beliefs evolve through natural-language interaction. Two competing campaign agents—a proponent arguing for a given claim and a challenger arguing against it—operate under equal, finite resource constraints while attempting to influence this population. The claims span false, contested, and well-supported positions, allowing us to examine how campaign outcomes depend on both the direction of influence and the nature of the claim. This creates a two-sided sequential influence problem in which outcomes depend jointly on resource-allocation strategy, language-mediated persuasion, population heterogeneity, and model-intrinsic opinion tendencies.

We use this setting to study two questions. 

\noindent \textbf{RQ1}: How do topic-dependent uninfluenced trajectories shape the apparent effect of competing campaigns in LLM populations?\\
\noindent\textbf{RQ2}: Do temporal resource-allocation effects observed under classical opinion dynamics persist when the recipients are LLM agents?



\section{Related Work}
\label{sec:related_work}

Our work sits at the intersection of LLM-based opinion dynamics, adversarial and corrective influence in LLM populations, and competitive influence games.

\subsection{LLM-Based Opinion Dynamics}

Recent work has begun to replace numerical opinion-update rules with natural-language interaction between LLM agents. Chuang et al.~\cite{chuang2024simulating} show that networked LLM populations tend to converge towards scientifically accurate consensus unless additional behavioural tendencies, such as confirmation bias, are introduced. Wang et al.~\cite{wang2025decoding} similarly study networked LLM agents under recommendation mechanisms and compare their collective behaviour with bounded-confidence and Friedkin--Johnsen models. Cau et al.~\cite{cau2025selective} remove explicit averaging rules altogether and show that repeated LLM interaction produces asymmetric persuasion and opinion convergence through language-mediated exchange. He et al.~\cite{he2026opinion} further connect LLM dialogue dynamics with Friedkin--Johnsen-style anchoring by retaining agents' initial opinions across rounds.

More recent studies show that these dynamics depend strongly on the underlying model. Probine et al.~\cite{probine2026characterizing} find that adding an intrinsic opinion-bias term substantially improves the fit of classical models to networked LLM behaviour. Brockers et al.~\cite{brockers2026disentangling} separate interaction effects from topic, agreement, and anchoring biases, showing that collective trajectories may be driven partly by model-specific default positions. This is directly relevant to strategic influence: a population may move towards one side because of persuasive interaction or because that side is already closer to the model's intrinsic attractor. These findings raise a further question for influence simulations, whether changes in average stance and changes in opinion diversity follow the same trajectory. A population may move towards a particular position while simultaneously becoming more homogeneous, making aggregate opinion shifts insufficient to characterise its collective dynamics.

These observations also motivate caution when interpreting LLM populations as proxies for humans. The DEBATE benchmark~\cite{chuang2025debate}, based on multi-round human discussions across more than one hundred topics, finds systematic differences between human and role-playing LLM group dynamics, including excessive convergence. Accordingly, we treat LLM populations as artificial societies whose collective behaviour is itself of interest rather than as direct substitutes for human populations.

\subsection{Misinformation and Adversarial Influence in LLM Populations}

Several studies use LLM-based social simulations to investigate misinformation susceptibility, propagation, and mitigation. MOSAIC~\cite{liu2025mosaic} evaluates third-party, community-based, and hybrid fact-checking strategies in persona-conditioned social networks. PANDORA~\cite{borah2026persuasion} studies demographic variation in misinformation susceptibility and reports group-level polarisation under multi-agent interaction. Lin et al.~\cite{lin2026cantfool} examine how traits such as open-mindedness and ideology affect community responses to misinformation shocks and corrective interventions. Dehkordi et al.~\cite{dehkordi2026polarization} show that a budget-limited adversary can increase polarisation in a network of persona-conditioned LLM agents and evaluate mechanisms for mitigating that effect.

Other work focuses on increasingly expressive simulation environments. Buitrago L\'opez et al.~\cite{buitrago2025agent} construct synthetic social networks with demographic and personality-conditioned agents, LLM-generated content, and a DISARM-inspired module for disinformation campaigns. Such frameworks provide rich environments for studying information operations but do not formulate misinformation and correction as a two-sided budget-constrained game. Moreover, when campaign messages are generated by LLMs, the intended direction of influence need not match the position actually expressed, creating an additional measurement challenge for evaluating campaign effectiveness.

Nasim et al.~\cite{nasim2025simulating} present a simulator in which
competing LLM agents broadcast influence and counter-influence messages
to a network whose members have numerical opinions and interact through
bounded-confidence dynamics. Qasmi et al.~\cite{qasmi2025competing}
study a resource-constrained misinformation--debunking game with LLM
influencers and a bounded-confidence population. They find that
aggressive early debunking can produce short-term gains but deplete
resources needed later. In both studies, recipients update numerical
opinions according to a prescribed rule. Whether the temporal result
persists when recipients interpret messages and revise beliefs through
natural-language interaction remains open.

\subsection{Competitive Influence Games}

Competitive influence has an established literature in network games and opinion dynamics. Dhamal et al.~\cite{dhamal2020twophase} study two competing camps that divide finite budgets across two phases of a Friedkin--Johnsen opinion process, revealing a trade-off between early and late investment. Etesami~\cite{etesami2022openloop} considers multiple players allocating resources across network nodes and campaign times under DeGroot dynamics, while Bastopcu et al.~\cite{bastopcu2025online} examine online and offline versions of dynamic influence maximisation under limited budgets. More recent work studies adversarial and corrective influence under different network topologies and equilibrium assumptions~\cite{bayiz2025topology}.

These models provide formal insights into budget allocation, timing, targeting, and strategic competition. However, influence is represented through scalar investments or abstract signals, and recipient opinions evolve according to predefined numerical rules. The semantic content of messages, persona-dependent responses, and model-specific opinion biases are therefore outside the model.

Existing work consequently leaves a gap between two research traditions. Classical influence games model two-sided strategic competition under resource constraints but rely on numerical opinion dynamics; LLM-based misinformation simulations model natural-language persuasion and heterogeneous populations but rarely place both misinformation and correction under comparable sequential resource constraints. We bridge these traditions by embedding competing, budget-constrained LLM campaign agents within a population that revises opinions through natural-language interaction, and examine whether the strategic timing patterns observed under numerical opinion dynamics persist in this setting.

\input{simulation_setup}

\input{results}
\input{discussion}

\section{Conclusions}


We studied competing, resource-constrained influence in persona-conditioned LLM populations while separating campaign-associated change from the population’s uninfluenced trajectory. Across eight claims, baseline drift was topic dependent, while the population consistently lost stance diversity. Campaign outcomes therefore cannot be interpreted from raw opinion change alone. On the reference topic, campaign effectiveness depended on the opponent’s timing policy, producing a best-response structure that differed qualitatively from a matched bounded-confidence population. We also found that generative campaign agents do not always pursue their assigned side, making side-fidelity checks necessary when LLMs themselves act as influencers. Together, these results show that both endogenous population dynamics and the behaviour of generative campaign agents must be accounted for when evaluating strategic influence in artificial LLM societies.

\balance
\bibliographystyle{ACM-Reference-Format} 
\bibliography{sample}

\clearpage


\input{appendix}


\end{document}

%% file: simulation_setup.tex
\section{Simulation Setup}
\label{sec:setup}

We model a non-cooperative game in which two campaign agents compete for
influence over a population of persona-conditioned LLM agents. The design
has three kinds of component: \emph{agents}, \emph{measurement models} and \emph{policies}.
There are three agent types: (i) the population, (ii) a \prop{}, which argues that the topic claim is true, and (iii) a \chal{}, which argues that it is false. Measurement models observe but never act.
\emph{Measurement models} are a stance probe that
converts written opinions into numerical stances, and a message rater
that scores campaign messages. Neither is visible to any agent and
neither feeds back into the simulation. The strategic object is the
\emph{policy}: a policy fixes how a campaign agent spends its budget over
time. Campaign agents are named for what they do, not for whether they
are right or wrong. The topic set spans false, contested and
well-supported claims (Section~\ref{sec:conditions}), so on some topics
the \chal{} argues the weaker position.

\subsection{Population and Network}
\label{sec:population}

We consider the population as a directed graph $G=(\mathcal{V},\mathcal{E})$ on
$N$ agents. Agent $i$ reads the posts of
$\mathcal{N}^{-}(i)=\{j:(j,i)\in\mathcal{E}\}$. Influence is directional
and an agent does not know who reads it. In the default \emph{follow}
graph each agent follows $\bar{d}$ others chosen uniformly, so audience
size varies as it does when users choose whose opinions to read.

To study segregated populations we also use \emph{planted-partition} networks
\cite{condon2001algorithms}. The agents are split into two equal blocks and each ordered pair of agents is an edge with probability $p_{\mathrm{in}}$ within a block and $p_{\mathrm{out}}$ between blocks. We parameterise the structure by the block ratio $\lambda_{\mathrm{r}}=p_{\mathrm{in}}/p_{\mathrm{out}}\ge1$, the factor by which within-block ties are more likely than between-block ties, with $p_{\mathrm{in}}$ then set so the mean degree equals $\bar{d}$. At $\lambda_{\mathrm{r}}=1$ this is a directed
Erd\H{o}s--R\'enyi graph and serves as the structureless control with
the same degree distribution. Asymptotic detectability
bounds~\cite{decelle2011asymptotic} are unreliable at $N=30$, so we verify
recoverability directly through spectral bisection that recovers the planted blocks
with accuracy $\ge0.8$ in $92\%$ of graphs at $\lambda_{\mathrm{r}}=4$
and $67\%$ at $\lambda_{\mathrm{r}}=3$. We therefore use
$\lambda_{\mathrm{r}}\ge4$ and record the realised within-block edge
fraction of every graph. Each agent follows at least one agent in its
own block.

Community structure and opinion segregation are varied separately. A
parameter $\alpha\in[0.5,1]$ sets the probability that an agent's
starting attitude is drawn from the pool matching its block, with the
marginal distribution of attitudes held fixed; $\alpha=0.5$ gives
structure without segregation and $\alpha=1$ gives fully opinion-sorted
blocks. Because more agents start rejecting than accepting, full
alignment produces one nearly homogeneous rejecting block and one mixed
block. The two $\alpha$ values share graphs, personas and drawn
attitudes and differ only in how attitudes are sorted into blocks.

\paragraph{Personas.}
Each agent holds a persona $p_i$. Each persona combines an age band, an
occupation, a location, a news habit and a writing style; occupations
carry a minimum age band so combinations remain plausible. Personas
carry no attribute relating to opinion or susceptibility. Starting
opinions are assigned separately (Section~\ref{sec:conditions}), and
anchoring and susceptibility are governed by the parameters $\lambda$
and $w$ of Section~\ref{sec:update}. Anchoring is the weight an agent
gives to its own starting opinion when it revises, so that revision is
not memoryless; susceptibility is the probability that an agent reads a
campaign message addressed to it.

\paragraph{Textual representation of opinions.}
An agent's belief state at round $t$ is a written \emph{opinion}
$s_i^{t}$ of at most $\ell$ tokens, rewritten each round; the same text
is what its followers read. Campaign agents seed the network, but the
population generates almost all of its language. For instance, one run at $N=30$
yields $1{,}020$ population opinions and at most $50$ campaign
messages. A paid message reaches its named recipients; what they write
next reaches everyone who follows them, and this indirect path carries
the volume.
Private belief and public utterance are not separated; one text serves
as both.

\paragraph{Starting attitudes are assigned.}
Each agent draws a target $x_i^{\star}$ from a distribution with a
rejecting group, a smaller accepting group and an undecided remainder,
following Qasmi et al.~\cite{qasmi2025competing}. The target is rendered as a
phrase (``you are sceptical of this claim'') and the agent writes
$s_i^{0}$ expressing that attitude in its own voice; it never sees the
number, and $s_i^{0}$ is retained as the anchor of
Section~\ref{sec:update}. We assign initial attitudes because preliminary experiments produced uniformly neutral opinions. When agents chose their own opening
opinion in a pilot, every persona hedged, every opinion read as
$0.00$, and polarisation was identically zero. It also makes the
starting distribution a controlled parameter rather than an artefact of
the population model. With attitudes assigned, the probe recovered them
at $r=0.97$ (mean absolute error $0.15$).

\subsection{Stance Measurement}
\label{sec:measurement}
Similar to traditional opinion dynamics models, the analysis needs numbers, but asking the agents for them would make the
population both the thing measured and the instrument. A \emph{stance
probe} $\Pi$, from a different model family than the population, run at
temperature zero and fixed across conditions, maps an opinion to
\begin{equation}
  x_i^{t} \;=\; \Pi\!\left(s_i^{t}\right) \;\in\; [-1,1],
  \label{eq:probe}
\end{equation}
where $-1$ is full rejection and $+1$ full acceptance of the claim.

No stance value exists during a run. Opinions are logged and the probe
is applied afterwards, so measurement adds nothing to a round and can
be redefined without re-simulating. Agents never see numeric values:
the population is instructed not to use numbers or scales, and campaign
agents are shown no stance, individual or aggregate. In production,
opinions are scored in batches of twenty with duplicate texts read
once; each batch carries two fixed anchor opinions of known stance, and
a batch whose anchors are misread is re-queried or marked missing.
Probe validity is checked in two ways, both reported in
Appendix~A.3: a second probe from a larger model of the
same family re-scores a stratified sample of 320 logged opinions
(Pearson $r{=}0.93$, Krippendorff's $\alpha{=}0.92$), and every
campaign message is classified, blind to its label, for the side it
actually argues, since a campaign agent that declines its brief would
otherwise be read as a campaign that failed. Production measurement
uses one reading per opinion.

\subsection{Campaign Agents and the Budget Game}
\label{sec:game}

Both campaign agents are LLM agents with role instructions, an equal
budget $B_{\mathrm{P}}=B_{\mathrm{C}}=B$, and identical prices.

\paragraph{Budget-constrained message delivery.}
At round $t$ side $c\in\{\mathrm{P},\mathrm{C}\}$ spends
$a_c^{t}\in\mathbb{Z}_{\ge0}$ and delivers its message to a set
$\mathcal{T}_c^{t}\subseteq\mathcal{V}$ of that size, subject to
\begin{equation}
  \sum\nolimits_{t=1}^{T} a_c^{t} \;\le\; B .
  \label{eq:budget}
\end{equation}
One unit of budget delivers one message to one agent. This is the
central departure from Qasmi et al.~\cite{qasmi2025competing}, where messages
were broadcast to the whole population in each round. When messages are broadcast we do not face an allocation problem, whereas paying per recipient makes the timing of
expenditure consequential. A round carries at most $N$ deliveries, so a
policy demanding more is capped at $N$ and the excess carried forward.
We set $B=\tfrac{1}{3}NT$, which keeps a concentrated policy close to
fully deliverable. Because the question concerns timing rather than
selection, $\mathcal{T}_c^{t}$ is drawn uniformly at random.

\paragraph{Timing policies.}
A policy spreads $B$ over a set $\mathcal{R}$ of active rounds, placing
the rounding remainder in the first. With $k=|\mathcal{R}|$,
$q=\lfloor B/k \rfloor$ and $r = B - kq$,
\begin{equation}
  \pi(t) =
  \begin{cases}
    q + r & t = \min \mathcal{R},\\
    q     & t \in \mathcal{R}\setminus\{\min \mathcal{R}\},\\
    0     & \text{otherwise},
  \end{cases}
  \label{eq:policy}
\end{equation}
so $\sum_t \pi(t)=B$ exactly. The three policies differ only in
$\mathcal{R}$: an \emph{early} policy spends in
$\{1,\dots,\lceil T/3 \rceil\}$, a \emph{sustained} policy in
$\{1,\dots,T\}$, and a \emph{late} policy in
$\{\lfloor 2T/3 \rfloor+1,\dots,T\}$, spanning the early-versus-late
trade-off identified by Dhamal et al.~\cite{dhamal2020twophase} for classical
opinion dynamics. Total exposure is identical across the three; only its
timing differs. The remainder is placed first because the late policy's
last active round is $T$, where an over-cap remainder would spill past
the campaign and be lost. The strategy space is the $3\times3$ product of policies,
treated as a normal-form game; we report the payoff matrix and its
pure-strategy equilibria.

\paragraph{What campaign agents observe.}
Expenditure is not an agent decision: $\pi_c$ is fixed before the run
and targets are random, because a normal-form game requires strategies
committed in advance, as in the two-phase investment game of Dhamal et
al.~\cite{dhamal2020twophase}. Message content, by contrast, adapts. In each
active round a campaign agent is shown the round index and the campaign
horizon $T$, its remaining budget and this round's reach, the number of
agents the other side reached in the previous round, its own previous
message, and a random sample of $n_{s}$ current population opinions, as
text. It is shown no stance value: conditioning on probe output would
place the probe inside every round and make the instrument part of the
mechanism it measures.

\paragraph{Rating campaign messages.}
A message rater $J$ scores every message on five dimensions in $[0,1]$: \emph{clarity}, \emph{evidential grounding}, \emph{relevance}, \emph{emotional appeal} and
\emph{persuasiveness}. Each dimension is elicited separately through a constrained output
schema. The dimensions are not combined. A false claim can be
persuasive and poorly grounded while a correct rebuttal is well grounded
and unpersuasive, and an average would assign both the same value. No
rating enters any update, so realised movement can be regressed on
measured quality rather than caused by it through construction.
The rater is validated once by scoring a sample $k_{J}$ times
and reporting per-dimension $\mathrm{ICC}(1,1)$.

\subsection{Round Structure and Belief Revision}
\label{sec:update}

A run comprises $T$ campaign rounds followed by $T'$ quiet rounds in
which budgets are exhausted and no messages are sent; the quiet rounds
distinguish a transient gain from a durable one.

Within a round, agent $i$ rewrites its opinion given its persona, its
current opinion, the current opinions of the agents it follows, and
any campaign message addressed to it. Neighbours' opinions arrive as
text and the agent reads all of them: unlike the bounded-confidence comparator of
BCM, there is no confidence threshold filtering distant views, and whether disagreement moves an
agent is left to the model. The starting opinion $s_i^{0}$ re-enters
with weight $\lambda$, providing resistance to persuasion in the manner
of Friedkin--Johnsen retention~\cite{friedkin1990social,he2026opinion}. The
probability $w$ that an addressed agent reads its message is fixed at
$1$. Updates are synchronous: every agent revises from the same
snapshot, so a change reaches followers in the following round.
Algorithm~\ref{alg:round} summarises one round.

\begin{algorithm}[t]
\caption{One round}
\label{alg:round}
\begin{algorithmic}[1]
\Require opinions $\{s_i^{t}\}$, budgets $\{B_c^{t}\}$, policies $\{\pi_c\}$
\For{$c \in \{\mathrm{P},\mathrm{C}\}$ \textbf{in parallel}}
  \State $a_c^{t} \gets \min(\pi_c(t)+\mathrm{carry}_c,\; B_c^{t},\; N)$
  \State $\mathrm{carry}_c \gets \min(\pi_c(t)+\mathrm{carry}_c, B_c^{t}) - a_c^{t}$
  \If{$a_c^{t} > 0$}
    \State $m_c^{t} \gets \textsc{Generate}(c, \text{claim}, \mathrm{ctx}_c^{t})$
    \State $\mathcal{T}_c^{t} \gets$ sample $a_c^{t}$ agents uniformly
    \State $B_c^{t+1} \gets B_c^{t} - a_c^{t}$
  \EndIf
\EndFor
\For{$i \in \mathcal{V}$ \textbf{in parallel}}
  \State $M_i \gets \{\, m_c^{t} : i \in \mathcal{T}_c^{t} \,\}$
  \State $s_i^{t+1} \gets \textsc{Revise}\big(p_i, s_i^{t},
          \{s_j^{t}\}_{j\in\mathcal{N}^{-}(i)}, M_i, s_i^{0}\!\mid\!\lambda\big)$
\EndFor
\State \textbf{log} $\{s_i^{t+1}\}$, $\{m_c^{t}\}$, $\{a_c^{t}\}$, $\{\mathcal{T}_c^{t}\}$
\Statex \Comment{stance and ratings are computed later, from the log}
\end{algorithmic}
\end{algorithm}

\subsection{Topics and Conditions}
\label{sec:conditions}

Topics are the unit of replication for the question of how the
population's untouched trajectory shapes apparent advantage, and must
therefore differ in unaided drift. Candidates
are screened by running each briefly with no campaign agents and sorting
by the direction and size of the resulting drift. We retain eight claims spanning accepting, undecided and rejecting
defaults (Table~\ref{tab:topics}). The claim with drift closest to zero
is the \emph{reference topic} and carries the timing experiment, so that
policy effects are measured where the population pulls least and both
sides have comparable room to move.

\input{topics_table}

Each topic is run under four conditions: \textsc{Null}, with no campaign
agents; \textsc{\Prop-only} and \textsc{\Chal-only}; and
\textsc{Competition}, both sides under a given policy pair.

\textsc{Null} serves two purposes. Its trajectory is itself the object of RQ1, since it is the uninfluenced, topic-dependent belief trajectory; and it is the baseline for every other condition, each reported effect being the difference from the \textsc{Null} run on the same topic, graph and personas.

One further check is run on the reference topic: the network is
replaced by the planted partition of Section~\ref{sec:population} in
three cells --- the structureless control ($\lambda_{\mathrm{r}}=1$),
communities without segregation ($\lambda_{\mathrm{r}}=4$,
$\alpha=0.5$) and segregated communities ($\lambda_{\mathrm{r}}=4$,
$\alpha=1$) --- so that the effect of community structure is separated
from that of opinion segregation. The segregated cell is
run on two further topics, vaccines and MMR, so that the within- and
between-community components of stance diversity can be compared across
a contested, a well-supported and a false claim. (At $\lambda_{\mathrm{r}}=1$ there
are no blocks to sort into, so $\alpha$ is undefined and the control is
a single cell.) All four conditions are run in each cell. As a comparator for RQ2, the
timing matrix is also replayed on a bounded-confidence
population~\cite{deffuant2000mixing} with the same graphs, starting attitudes
and deliveries; its update rule and parameters are given in
Appendix~A.8. The four corner cells of the matrix are
additionally run on vaccines and on the segregated planted network
(Appendix~A.7).

\subsection{Measures and Data Record}
\label{sec:measures}

With $\bar{x}^{t}=N^{-1}\sum_i x_i^{t}$ we report polarisation
$P^{t}=N^{-1}\sum_i (x_i^{t}-\bar{x}^{t})^{2}$, the measure of Chitra
and Musco~\cite{chitra2020analyzing}; advantage
$A^{t}=\bar{x}^{t}-\bar{x}^{t}_{\textsc{Null}}$, positive when the
\prop{} is ahead; drift-corrected polarisation $\Delta P^{t} = P^{t} -
P^{t}_{\textsc{Null}}$; the shares of agents rejecting, undecided and
accepting ($x\le-\tau$, $|x|<\tau$, $x\ge\tau$); and persistence
$\rho_{\mathrm{pers}} = A^{T+T'}/A^{T}$, the fraction of the
end-of-campaign advantage remaining after the quiet rounds. An opinion
the probe cannot read is a missing observation, not a zero, and is
excluded from all statistics.

The unit of replication is the run. Each condition is repeated $R$
times; replicate $r$ of every condition and every topic uses the same
graph, personas and starting attitudes, so all contrasts are paired by
replicate and only the campaign differs within a pair. Every quantity is a mean
over runs with a bootstrapped $95\%$ interval, normal-theory intervals
being inappropriate for the deliberately spread starting distribution.
The smallest detectable difference is reported with the results.

All generated opinions and campaign messages, including their recipients, are retained and exported as topic-specific datasets alongside the corresponding network structures. This enables stance measurements to be repeated using alternative probes without rerunning the simulations.

\subsection{Configuration}
\label{sec:config}

Table~\ref{tab:config} lists the parameters with the values used;
those marked $\dagger$ were fixed before any experiment run and never
varied. Table~\ref{tab:blocks} lists the experimental blocks and their
replication. The four per-topic blocks and the timing block share one
set of twenty networks, personas and starting attitudes, so every
contrast within a topic is a paired comparison on the same population;
the network-structure check uses its own graphs and is compared only
within its generator.

\begin{table}[t]
\centering\small
\caption{Parameters. Values marked $\dagger$ are held fixed. Four models
are used: population, campaign generation, message rating and stance
probe; the probe is from a different family than the population.}
\label{tab:config}
\begin{tabular}{llr@{\hskip 10pt}llr}
\toprule
Sym. & Meaning & Val. & Sym. & Meaning & Val. \\
\midrule
$N$ & agents & 30 & $\tau$ & band cutoff & 0.5 \\
$\bar{d}$ & agents followed & 4 & $\lambda^{\dagger}$ & anchor weight & 0.3 \\
$T$ & campaign rounds & 25 & $w^{\dagger}$ & read prob. & 1.0 \\
$T'$ & quiet rounds & 8 & $n_{s}^{\dagger}$ & posts shown & 5 \\
$B$ & budget & $\tfrac{NT}{3}$ & $k_{J}^{\dagger}$ & rater repeats & 5 \\
$\ell$ & opinion tokens & 80 & $\lambda_{\mathrm{r}}$ & block ratio & $\{1,4\}$ \\
$n_{\mathrm{top}}$ & topics & 8 & $\alpha$ & alignment & $\{0.5,1\}$ \\
\bottomrule
\end{tabular}
\end{table}

\begin{table}[t]
\centering\small
\caption{Experimental blocks. The sustained--sustained competition cell is
shared between the timing experiment and the per-topic condition set at
the reference topic; the eight remaining policy pairs are the timing
block. The network-structure check runs all four conditions in each of
three planted-partition cells on the reference topic, and in the
segregated cell on two further topics.}
\label{tab:blocks}
\begin{tabular}{lrrr}
\toprule
Block & Cells & $R$ & Runs \\
\midrule
\textsc{Null}, per topic          & 8          & 20 & 160 \\
\textsc{\Prop-only}, per topic    & 8          & 16 & 128 \\
\textsc{\Chal-only}, per topic    & 8          & 16 & 128 \\
\textsc{Competition}, per topic   & 8          & 16 & 128 \\
Timing policies, reference topic  & 8          & 8  & 64 \\
Network-structure check           & $3\times4$ & 16 & 192 \\
Timing corner cells, vaccines     & 4          & 8  & 32 \\
Timing corner cells, segregated network & 4    & 8  & 32 \\
Segregated cell, vaccines and MMR & $2\times4$ & 16 & 128 \\
\midrule
\multicolumn{3}{l}{Total}                      & 992 \\
\bottomrule
\end{tabular}
\end{table}
The simulation used Claude Sonnet 5 for campaign agents, Claude Haiku 4.5 for stance measurement, and GPT-5-nano for message evaluation. Population agents shared a common underlying LLM, with distinct personas and opinion histories.







%% file: topics_table.tex
\begin{table*}[t]
\centering\small
\caption{The eight topic claims. The \prop{} argues for the claim and the \chal{} against it
in every condition. Claims are chosen to
cover the range of the population model's default position, from ones it
rejects outright to ones it accepts, so that the untouched trajectory
varies across the set instead of sitting at one extreme on every topic.
\texttt{decouple} is the reference topic for the timing experiment.}
\label{tab:topics}
\begin{tabular}{lllr}
\toprule
ID & Claim (the \prop{} argues for it) & Status & Default stance \\
\midrule
\texttt{mmr} & Childhood MMR vaccination is a cause of autism & clearly false & $-0.99$ \\
\texttt{5g} & 5G mobile network infrastructure causes adverse health effects in humans & clearly false & $-0.85$ \\
\texttt{gmo} & Genetically modified foods approved for sale pose distinctive risks to human health & mostly false & $-0.75$ \\
\texttt{remote} & Remote work reduces overall worker productivity & contested & $-0.37$ \\
\texttt{socmed} & Social media use is a principal cause of the recent rise in adolescent mental-health problems & contested & $-0.16$ \\
\texttt{decouple} & Economic growth can be fully decoupled from carbon emissions & contested & $-0.05$ \\
\texttt{nuclear} & Nuclear power is the safest large-scale electricity source per unit of energy delivered & well supported & $+0.20$ \\
\texttt{vaxmort} & Vaccination programmes have substantially reduced childhood mortality worldwide & well supported & $+0.89$ \\
\bottomrule
\end{tabular}
\end{table*}

%% file: results.tex
\section{Results}
\label{sec:results}
\providecommand{\mainfig}[2]{\includegraphics[width=#1]{#2}}

\begin{figure}[t]
\centering
\includegraphics[width=0.95\linewidth]{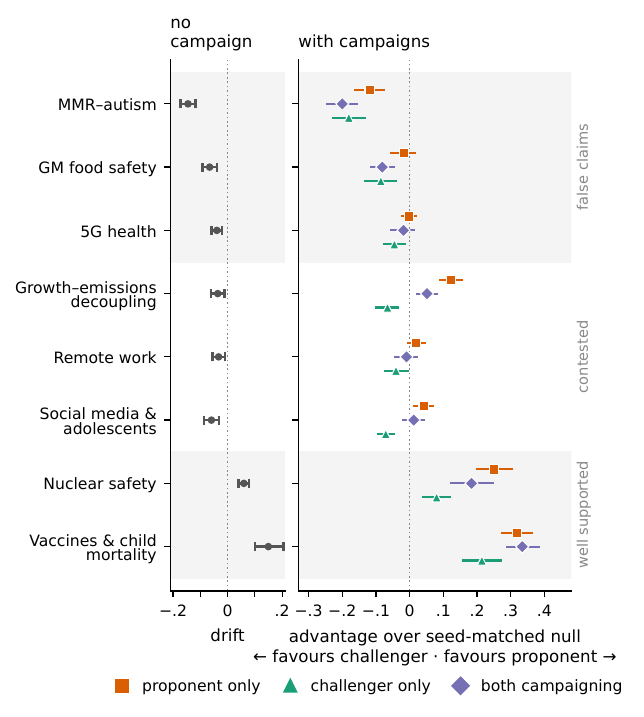}
\caption{Eight topics ordered by where the uninfluenced population
ends, from the claim most rejected to the claim most accepted. Left:
drift of the no-campaign population over the run (final mean stance
minus starting mean stance; $n{=}20$ runs). Right: advantage of each
campaign condition after the eight quiet rounds, the population's mean
stance minus that of the no-campaign run on the same network, personas
and starting attitudes; positive favours the \prop{} ($n{=}16$ per
condition). Points are means with 95\% bootstrap CIs over runs.}
\label{fig:cross_topic}
\end{figure}
\subsection{RQ1: the uninfluenced trajectory and the apparent advantage}

In the absence of campaign interventions, the population exhibits topic-dependent shifts in mean stance (Figure~\ref{fig:cross_topic}, left). We define baseline drift as the change in population mean stance between the initial and final rounds of a no-campaign simulation. Across the eight claims, this drift ranges from $-0.15$ for MMR to $+0.15$ for vaccination and childhood mortality. The direction follows the epistemic standing of the
claim rather than the starting distribution, which is identical across
topics: populations drift toward rejecting false claims and toward
accepting well-supported ones, and move least, though still toward
rejection, on contested ones. Any
comparison of campaign conditions that ignores this drift attributes it
to whichever side is being measured; on MMR, drift accounts for roughly
two fifths of the raw change in mean stance under competition. The drift is directional only: the
uninfluenced population also loses $29$--$45\%$ of its stance variance
on every topic, by an amount unrelated to the sign or size of its drift
(Appendix~A.5).

\begin{figure*}[t]
\centering
\includegraphics[width=0.8\linewidth]{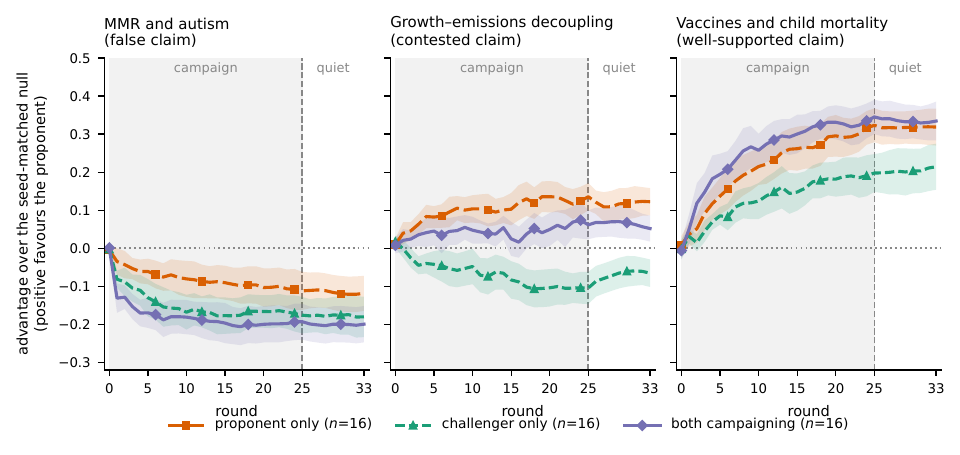}
\caption{Advantage over the seed-matched no-campaign run, round by
round, for one topic of each kind: MMR--autism (false), growth--emissions
decoupling (contested) and vaccination and child mortality (well
supported). Each line is the mean over runs of the population's mean
stance under that condition minus the no-campaign run on the same
network; positive favours the \prop{}; bands are 95\% CIs across runs.
Campaigns run in rounds 1--25 (shaded); rounds 26--33 are quiet. In none
of the three do the quiet rounds pull the effect back toward zero.}
\label{fig:exemplars}
\end{figure*}

Once the drift is removed by differencing against the seed-matched
null, a campaign's effect also follows the claim rather than the side
that speaks (Figure~\ref{fig:cross_topic}, right). On the three false
claims every campaign condition sits at or below zero; on the two
well-supported claims every condition sits above zero, including the
\chal{} arguing against the claim ($+0.08$ on nuclear safety, $+0.21$
on vaccines); only on the three contested claims do the two sides move
the population in opposite directions, and there the contest lands
between them or near zero. Figure~\ref{fig:exemplars} shows how these
endpoints arise: on MMR the movement is complete within five rounds and
holds; on decoupling the two sides separate steadily and the contest
settles between them; on vaccines all three conditions climb for the
full campaign and are still rising when it stops. In none of the eight
topics does the eight-round quiet period return the effect toward zero.

Part of this pattern is produced by the campaign agents themselves, not
by the population. Classifying every campaign message for the side it
actually argues (Appendix~A.3) shows that the generation model
declined its brief exactly where the brief ran against the evidence.
On MMR, $97\%$ of the \prop{}'s messages argue that the claim is false
--- typically opening ``I will not argue that MMR causes autism'' and
then debunking it --- so the \prop{} acted as a second \chal{}, and the
MMR row of Figure~\ref{fig:cross_topic} measures two debunking
campaigns, not a contest. On 5G the \prop{} asserted the claim in only
$7\%$ of messages and hedged in $55\%$. In the other direction, the
\chal{} argued \emph{for} the claim in $58\%$ of its messages on
vaccines and $22\%$ on nuclear safety, which accounts for much of the
\chal{}'s positive effect there. On the three contested claims and on
GM food both sides argued their briefs in $76$--$100\%$ of messages;
the three contested claims are also the only topics on which the
single-sided effects have opposite signs. Claim semantics enter the simulation at two points: through the population’s topic-dependent trajectory and through whether the campaign model follows its assigned side. The stance scores themselves are not the
source of these patterns: a second probe from a different model
reproduces them closely. The RQ1
experiment is unaffected, since both sides argued their briefs on the
reference topic ($76\%$ and $100\%$; the remainder hedged rather than
switched).

\subsection{RQ2: timing}

\begin{figure}[tb]
    \centering
   \includegraphics[width=1\linewidth]{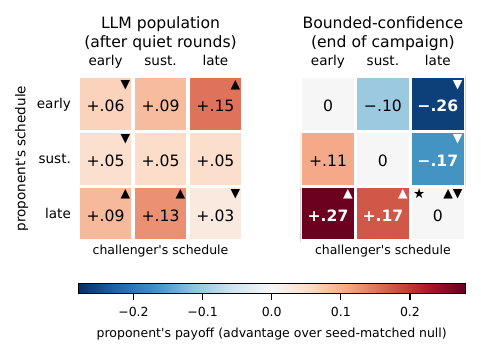}
\caption{Timing payoff matrices on the reference topic (growth--emissions
decoupling). All nine cells are two-sided contests with equal budgets
of 250 deliveries; cells differ only in each side's spending schedule
--- \emph{early} (rounds 1--9), \emph{sustained} (1--25) or \emph{late}
(17--25). Rows give the \prop{}'s schedule, columns the \chal{}'s; each
cell is the \prop{}'s payoff, the population's mean stance minus that
of the no-campaign run on the same network, personas and starting
attitudes, averaged over runs ($n{=}8$ per cell, $16$ for
sustained--sustained). \textbf{Left}: the LLM population, payoff after
the eight quiet rounds; 95\% bootstrap CI half-widths are
$0.03$--$0.10$ (full table in the appendix). \textbf{Right}: a
bounded-confidence population replayed on the same graphs, starting
attitudes and deliveries, payoff at the end of
the campaign, since its retention term erases campaign effects during
the quiet rounds. $\blacktriangle$ marks the \prop{}'s best reply in a
column, $\blacktriangledown$ the \chal{}'s in a row. In the LLM
population no cell carries both --- the \prop{} gains when the
schedules do not overlap, the \chal{} when they coincide --- so there is
no pure-strategy equilibrium (absent in 81\% of resamples). The
numerical population has exactly one, at late--late ($\bigstar$).}
\label{fig:payoff}
\end{figure}

With the drift removed, the timing experiment on the reference topic
yields the payoff matrix of Figure~\ref{fig:payoff}. Campaign outcomes depend on the interaction between the two expenditure schedules rather than on either schedule independently. The \prop{}'s payoff is largest when the two
sides do not spend in the same rounds (early against late, $+0.15$;
late against sustained, $+0.13$) and smallest when they do (early
against early, $+0.06$; late against late, $+0.03$); the mean
off-diagonal payoff exceeds the mean on-diagonal payoff by $0.042$
$[-0.001, 0.072]$ within matched seeds, a difference whose interval
just reaches zero. Consequently the \prop{}'s best
reply is always to avoid the \chal{}'s schedule, and the \chal{}'s is
to coincide with the \prop{}'s on the early and late schedules (against
a sustained \prop{} its three replies are indistinguishable). The estimated payoff matrix has no pure-strategy equilibrium, with this result reproduced in 81\% of bootstrap resamples. The sustained schedule, the only one studied in Qasmi et
al.~\cite{qasmi2025competing}, is never a best reply for either side.

The same schedules replayed on a bounded-confidence population --- the
same graphs, starting attitudes and deliveries, but numerical agents
(Appendix~A.8) --- give a different game
(Figure~\ref{fig:payoff}, right): late--late is a unique pure-strategy
equilibrium, present in $100\%$ of bootstrap resamples and unchanged
across $\lambda\in\{0,0.1,0.3\}$ and $\varepsilon\in\{0.3,0.6,1.0\}$. 
Under symmetric, memoryless interventions, later interventions have a systematic advantage. The absence of a stable schedule therefore, arises in the LLM population rather than from the allocation problem alone.

The best-reply structure is not specific to the reference topic or to
the follow graph. Repeating the four corner cells (early and late for
each side, $n{=}8$ per cell) on vaccines, a well-supported claim with
strong positive drift, and on the segregated planted network 
gives the same pattern in both. 
The
\prop{}'s best reply is to avoid the \chal{}'s schedule, the \chal{}'s
is to coincide with it, no cell is a best reply for both, and the
\prop{} gains more when the schedules do not overlap (paired
off-diagonal minus on-diagonal payoff $+0.09$ $[+0.03,+0.15]$ on
vaccines and $+0.07$ $[+0.00,+0.14]$ on the segregated network, against
$+0.08$ $[+0.01,+0.14]$ for the same four cells on the reference). The
absolute payoffs differ by topic --- every cell is above $+0.33$ on
vaccines, where both campaigns raise acceptance --- but the relation
between the two schedules is what decides the \prop{}'s advantage in
all three settings. These checks cover the four corner cells on one
additional topic and one additional network; they support the
best-reply structure as a property of this class of population rather
than a general timing rule. On the segregated network, campaigns primarily reduce between-community stance variance across the three tested topics, while within-community variance changes comparatively little (Appendix~A.4). 

%% file: discussion.tex
\section{Discussion}
Our results expose two features of LLM-based influence simulations that are obscured when campaign direction is treated as an externally controlled variable. First, the semantic and epistemic status of a claim enters the simulation at multiple points. The uninfluenced population follows topic-dependent trajectories, while the campaign model sometimes declines to advance an assigned position that conflicts strongly with available evidence. On MMR, for example, the nominal proponent overwhelmingly generated debunking messages instead. Consequently, an LLM campaign agent is not equivalent to a freely specified positive or negative intervention. Studies using generative agents for strategic influence should therefore verify side fidelity—whether generated messages actually pursue the assigned objective—rather than infer campaign behaviour from role labels alone.
Second, replacing numerical recipients with language-model agents changes the temporal game. In the bounded-confidence comparator, symmetric, memoryless interventions produce the expected advantage for speaking last and a stable late–late equilibrium. The LLM population instead produces cyclic best responses: the proponent benefits from avoiding the challenger’s timing, whereas for concentrated early or late schedules the challenger benefits from matching it. This pattern persists in the additional topic and network checks. The contrast suggests that the timing structure does not arise from budget allocation alone; nor does it arise from retention of initial positions by itself, since the numerical equilibrium is unchanged when that term is removed. Agents repeatedly reinterpret natural-language messages while retaining their initial positions, producing path-dependent and non-additive responses for which no single schedule is uniformly preferred.\\
The uninfluenced runs also reveal a distinction between direction and diversity. Mean stance drifts in different directions across claims, but stance variance falls on every topic. Campaigning further compresses this variance relative to matched no-campaign trajectories, regardless of the direction in which mean stance moves, and in segregated populations this compression falls almost entirely on the variance between communities rather than within them. 
Thus, directional stance change and population homogenisation are separable properties of these simulated societies; population averages alone can conceal substantial changes in their internal distribution.\\


These findings remain conditional on the tested simulation setting. The timing results cover two topics and two network settings, while community-level findings are limited to three topics on a single network family. The population comprises only 30 agents, restricting sensitivity to relatively large effects. Furthermore, agents express their beliefs through a single public opinion, preventing the simulation of strategic silence or differences between private and expressed beliefs. Campaign schedules are predetermined and recipients selected randomly, excluding adaptive budget allocation and targeted influence. Although stance measurements show strong cross-model agreement, they have not been validated against human judgements. Our findings therefore characterise artificial LLM populations rather than human opinion dynamics.

%% file: appendix.tex
\graphicspath{{figures-app/}}
\newcommand{\appfig}[2]{\includegraphics[width=#1]{#2}}

\renewcommand{\topfraction}{0.95}
\renewcommand{\dbltopfraction}{0.95}
\renewcommand{\textfraction}{0.05}
\renewcommand{\floatpagefraction}{0.85}
\renewcommand{\dblfloatpagefraction}{0.85}
\setcounter{topnumber}{4}
\setcounter{dbltopnumber}{4}
\setcounter{totalnumber}{6}


\appendix
\section{Supplementary Material}
\label{app:figures}
\setcounter{figure}{3}
\begin{figure}[b]\centering
\appfig{0.8\columnwidth}{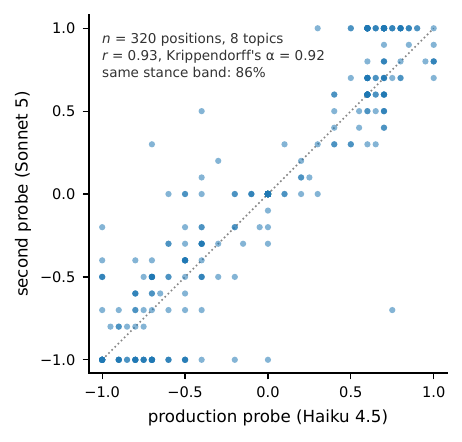}
\caption{Stance assigned by the production probe (Haiku~4.5) against a
second probe from a larger model (Sonnet~5) on 320 written positions,
40 per topic, stratified by stance band.}
\label{fig:probe_agreement}
\end{figure}
\subsection{The shared population}
\label{app:networks}
Replicate $r$ of every condition and every topic runs on network $r$
with the same personas and starting attitudes, so every contrast in the
paper is paired at the population level. Reading is fixed at four
sources per agent; audience is not (coefficient of variation
$0.37$--$0.54$), and six of the 600 agents have no readers. Attitudes are
assigned independently of position in the graph, so these populations
have structure without opinion segregation; segregation is introduced
only in the planted-partition check (Appendix~\ref{app:planted}).

\begin{figure*}[tp]
\centering
\appfig{\textwidth}{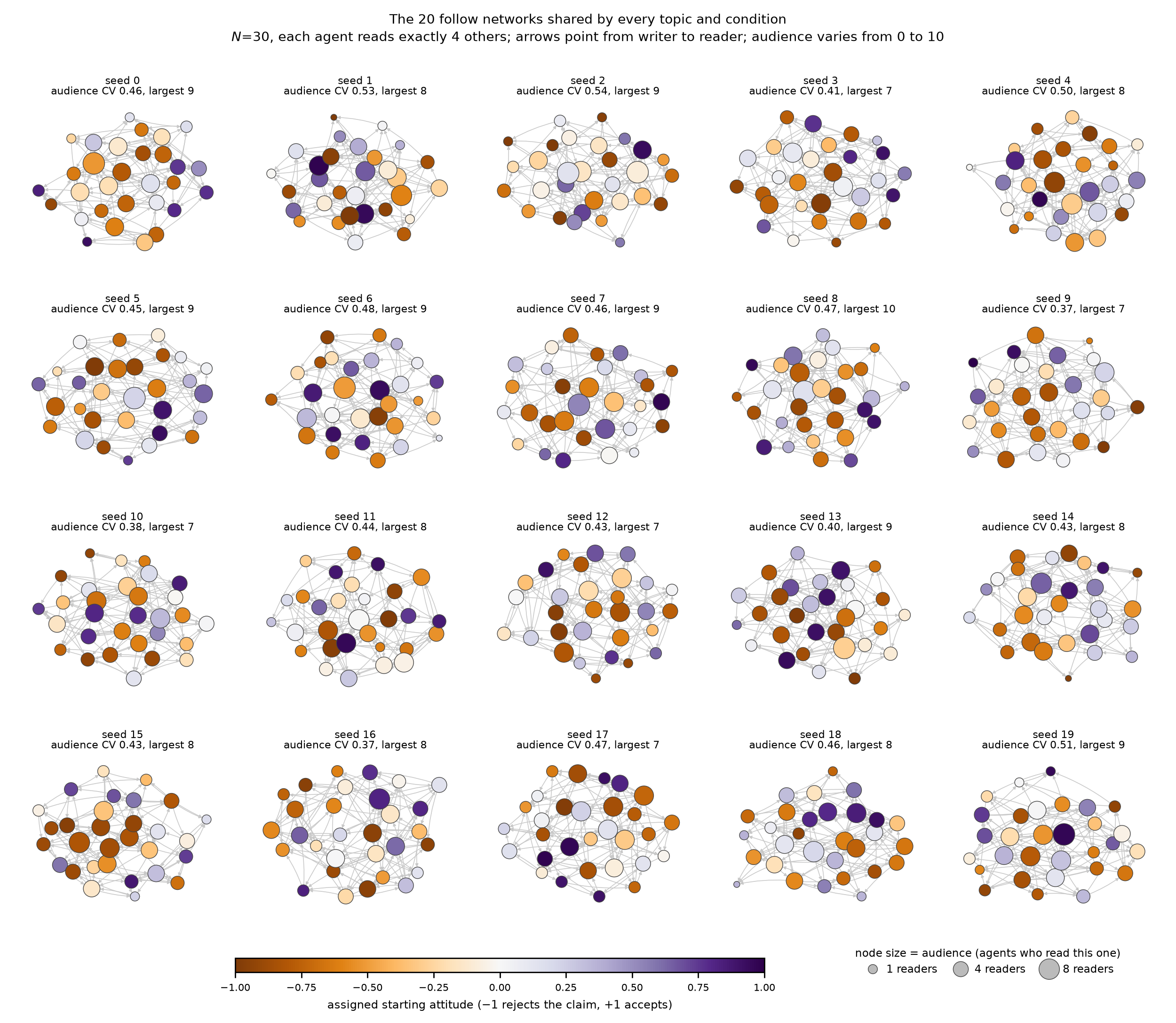}
\caption{The 20 follow networks used by every topic and condition.
$N{=}30$; each agent reads exactly four others, so arrows point from
writer to reader. Node colour is the assigned starting attitude, node
size the number of agents who read that node (0--10).}
\label{fig:networks}
\end{figure*}

\subsection{Per-topic trajectories}
\label{app:topics}

\begin{figure*}[tp]\centering
\appfig{0.7\textwidth}{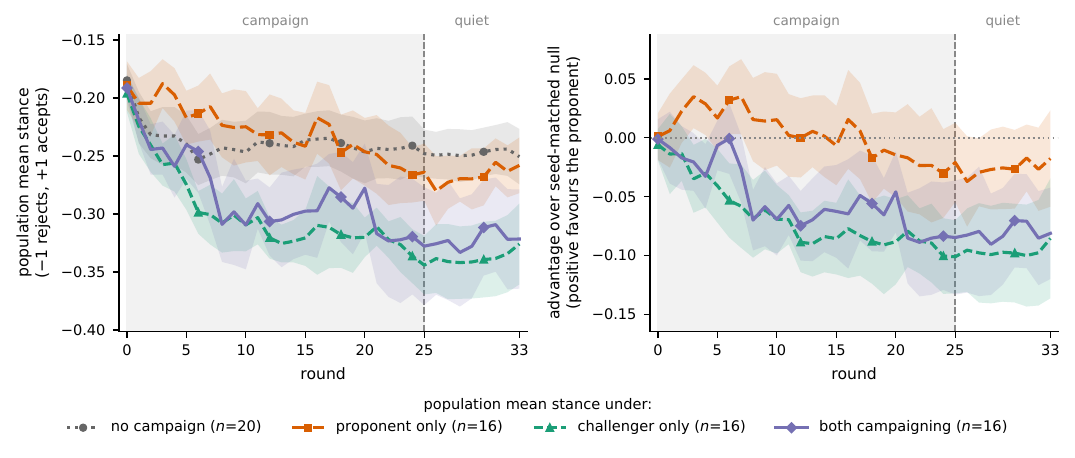}
\caption{GM food safety (mostly false). The \prop{}'s advantage peaks
near round~4 and returns to zero in the quiet rounds; the \chal{}'s
persists.}
\label{fig:t_gmo}
\end{figure*}

\begin{figure*}[tp]\centering
\appfig{0.7\textwidth}{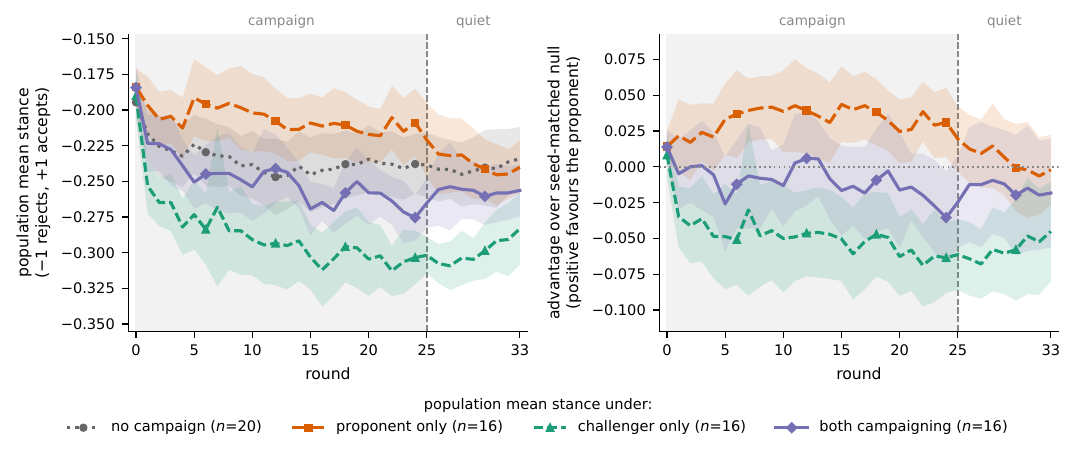}
\caption{5G health effects (false). As for GM food: a transient
\prop{} lift, a durable \chal{} effect, and a contest indistinguishable
from the null.}
\label{fig:t_5g}
\end{figure*}

\begin{figure*}[tp]\centering
\appfig{0.7\textwidth}{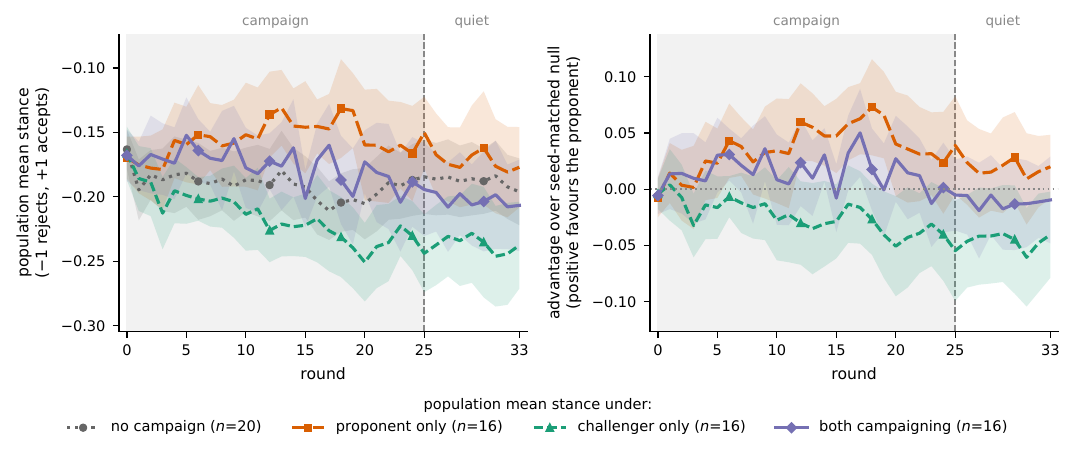}
\caption{Remote-work productivity (contested). The least movable topic:
all campaign effects are within $\pm0.05$ and the contest is zero.}
\label{fig:t_remote}
\end{figure*}

\begin{figure*}[tp]\centering
\appfig{0.7\textwidth}{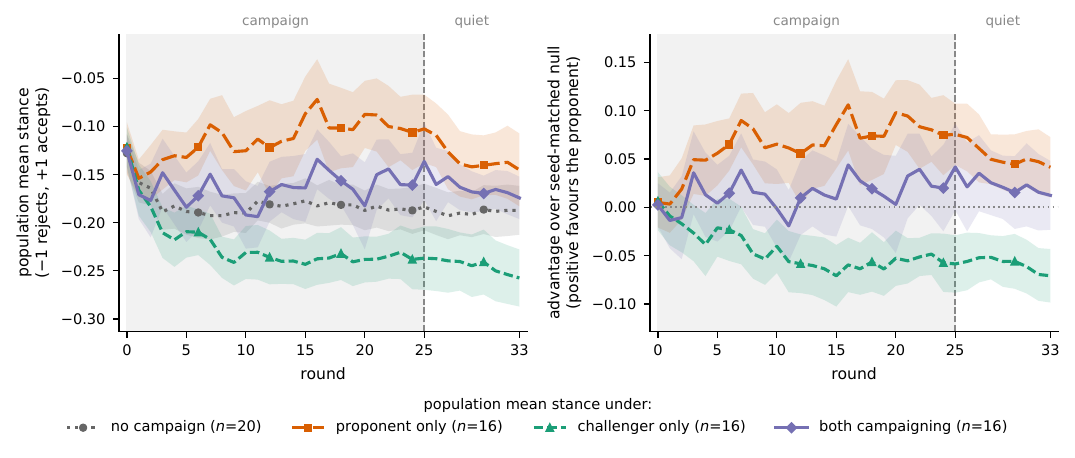}
\caption{Social media and adolescents (contested). Symmetric single-sided
effects of about $\pm0.05$ that cancel almost exactly in competition.}
\label{fig:t_socmed}
\end{figure*}

\begin{figure*}[tp]\centering
\appfig{0.7\textwidth}{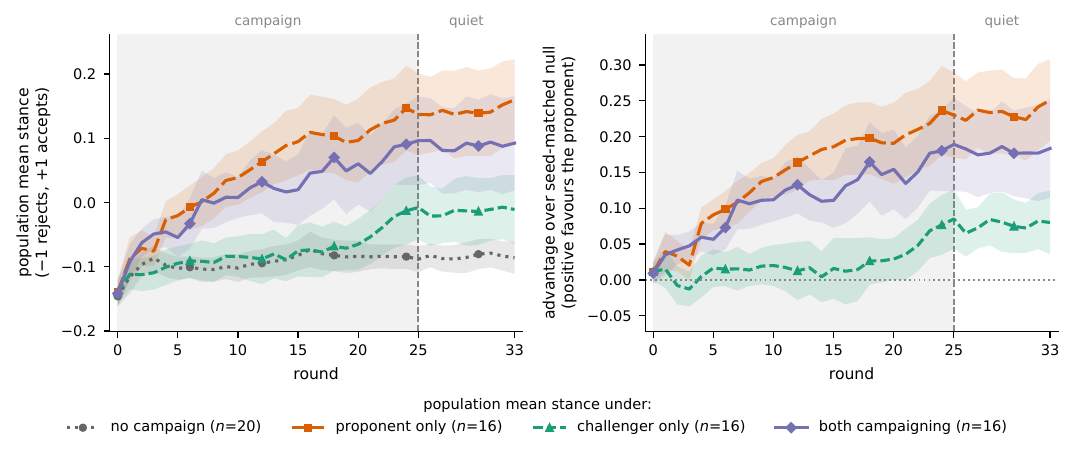}
\caption{Nuclear safety (well supported). Both campaigns raise
acceptance; the \chal{}'s effect appears only late in the campaign and
holds through the quiet rounds.}
\label{fig:t_nuclear}
\end{figure*}

Figures~\ref{fig:t_gmo}--\ref{fig:t_nuclear} give the five topics not
shown in Figure~2 of the main text, in the same two-panel form: the
left panel is the population's mean stance under each condition, the
right panel each campaign condition minus the no-campaign run on the
same network ($n{=}20$ no campaign, $n{=}16$ otherwise; bands are 95\%
CIs over runs). On the two false claims the \prop{} buys a small lift
during the campaign that does not survive the quiet rounds, while the
\chal{}'s effect persists. On the contested claims the two sides move
the population in opposite directions and the contest sits between
them; remote work is the topic the population barely moves on. On
nuclear safety both sides raise acceptance, the \chal{} included.

\subsection{Measurement validity}
\label{app:validity}

\begin{figure*}[tp]\centering
\appfig{0.8\textwidth}{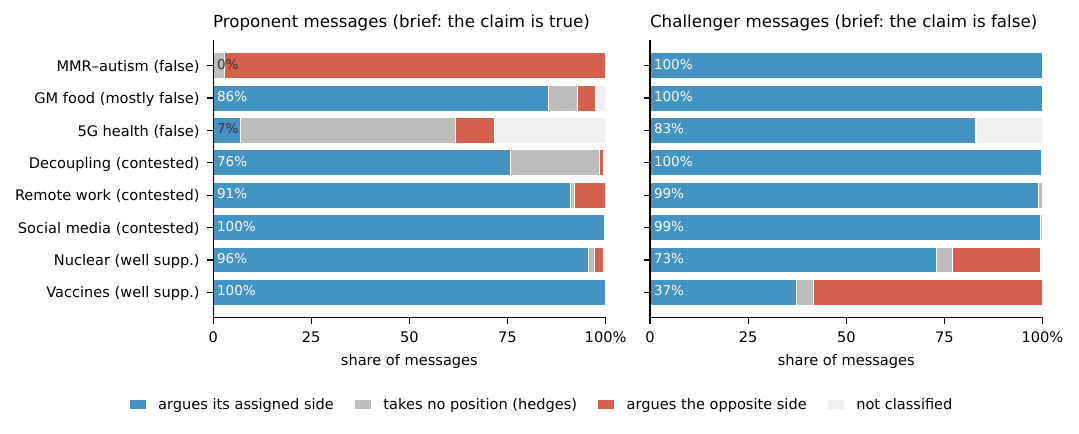}
\caption{Which side each campaign message actually argues, by topic
($n{=}768$--$1{,}632$ messages per bar). An independent model classified
every message, blind to its label, as arguing the claim is true, false,
or taking no position; each batch carried two anchor sentences of known
polarity and was discarded if either came back inverted (``not
classified''). The \prop{}'s brief is that the claim is true, the
\chal{}'s that it is false.}
\label{fig:side_fidelity}
\end{figure*}

\paragraph{Side fidelity.} Figure~\ref{fig:side_fidelity} shows that
the generation model argued its brief on the contested claims and on
GM food ($76$--$100\%$ of messages for both sides) but declined it where
the brief ran against the evidence: on MMR $97\%$ of \prop{} messages
argue the claim is false, on 5G the \prop{} asserted the claim in $7\%$
of messages and hedged in $55\%$, and the \chal{} argued for the claim
in $58\%$ of messages on vaccines and $22\%$ on nuclear safety. This is
a property of LLM campaign agents and is reported as a finding
(Section~4); the RQ2 experiment is unaffected because
both sides argued their briefs on the reference topic.


\paragraph{Probe agreement.} A second probe from a different model
re-scored 320 positions with the same prompt
(Figure~\ref{fig:probe_agreement}): Pearson $r{=}0.93$, Krippendorff's
$\alpha{=}0.92$ (interval), $86\%$ in the same stance band, four sign
disagreements, and a mean offset of $+0.04$ (Sonnet reads slightly more
accepting). This is cross-model agreement, not human validation; human
readings remain future work.

\clearpage
\begin{figure}[t]\centering
\appfig{0.8\columnwidth}{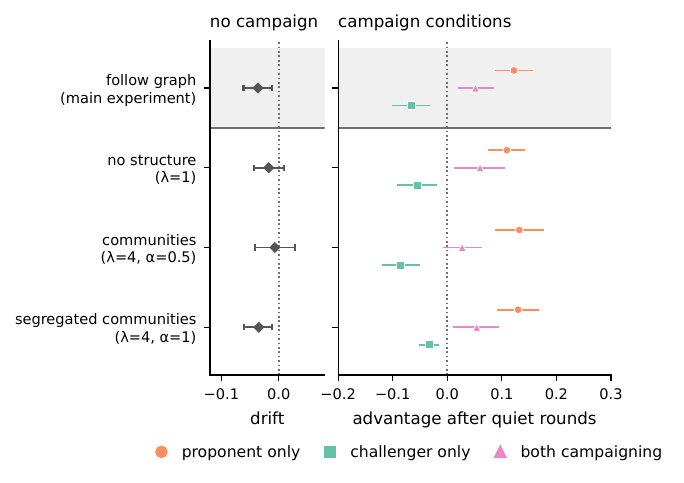}
\caption{Reference topic, four network cells. Top row: the follow graph
of the main experiment (reference). Below: planted-partition cells ---
no structure ($\lambda_{\mathrm r}{=}1$), two communities with
unsorted attitudes ($\lambda_{\mathrm r}{=}4$, $\alpha{=}0.5$), and
two opinion-sorted communities ($\alpha{=}1$). Left: drift of the
uninfluenced population; right: advantage after the quiet rounds,
positive favouring the \prop{}. 95\% bootstrap CIs; $n{=}16$ per
condition in every cell.}
\label{fig:planted_cells}
\end{figure}
\subsection{Community structure and opinion segregation}
\label{app:planted}

\begin{figure*}[tp]\centering
\appfig{0.7\textwidth}{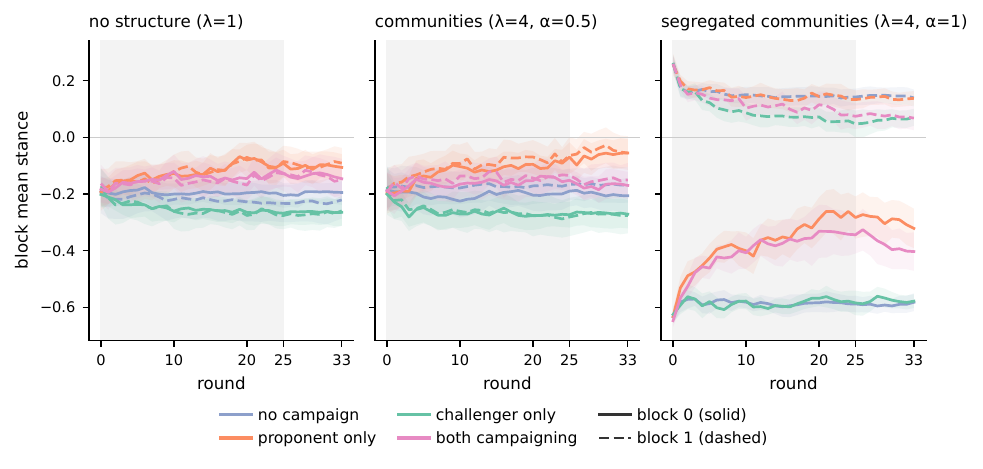}
\caption{Mean stance of each block separately (solid: block~0; dashed:
block~1) under each condition, for the three planted cells. In the
unsorted cells the two blocks are statistically identical. In the
segregated cell they start $0.89$ apart; the \prop{} lifts the rejecting
block by about $0.25$, while the \chal{} barely moves the accepting
block.}
\label{fig:planted_blocks}
\end{figure*}

The structureless control reproduces the follow-graph result, so the
change of generator itself does nothing. Adding communities leaves the
\prop{}'s effect unchanged; the one magnitude change is the \chal{}'s
effect, which halves under segregation ($-0.087$ to $-0.032$,
non-overlapping CIs). Figure~\ref{fig:planted_blocks} shows why: the
block that already agrees with the \chal{} has no room to move, and the
block it would need to reach is the one it does not. The asymmetry
reflects the skewed starting distribution (more rejecters than
accepters), which under full alignment yields one near-homogeneous
rejecting block and one mixed block.

\begin{figure*}[tp]\centering
\appfig{0.7\textwidth}{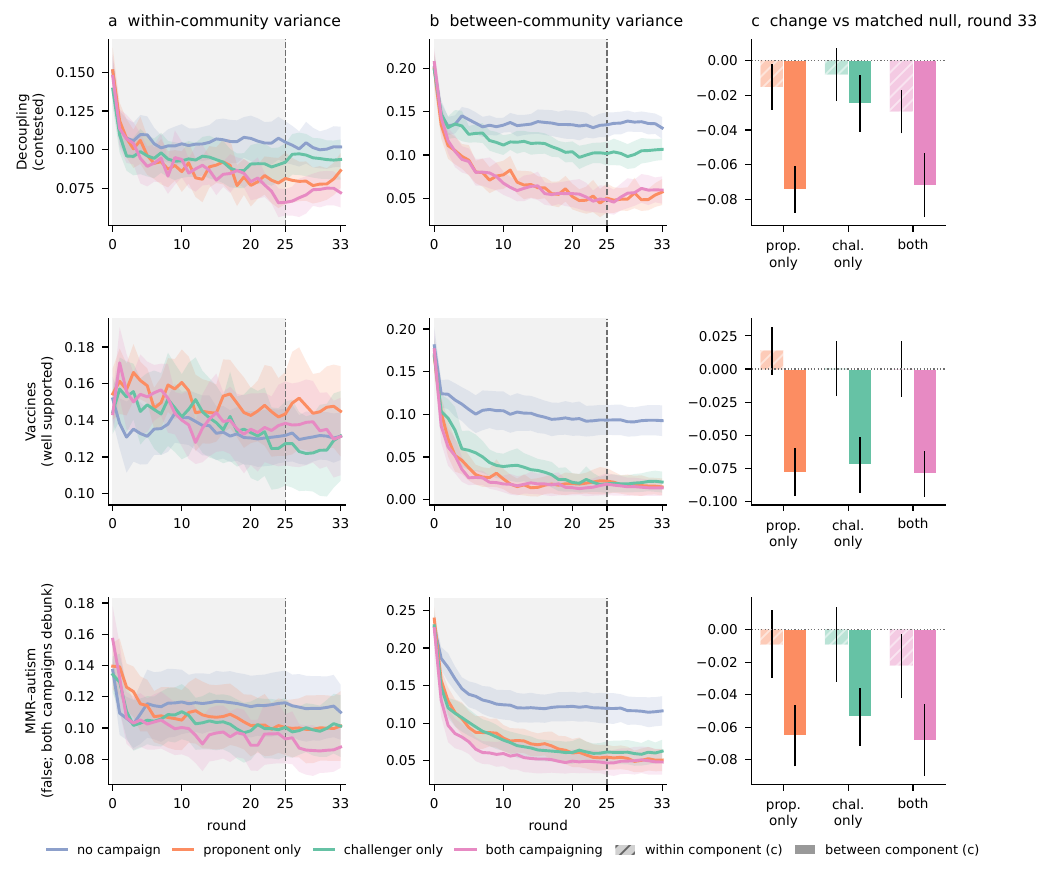}
\caption{Within- and between-community stance variance in the
segregated planted cell on three topics (rows), by condition; means
over $n{=}16$ runs with 95\% CIs. (a)~Within-community component by
round. (b)~Between-community component by round. (c)~Change in each
component at round~33 relative to the no-campaign run on the same
graph, personas and starting attitudes; bars are means with 95\%
bootstrap CIs over the sixteen matched pairs.}
\label{fig:within_between}
\end{figure*}

\paragraph{Within- and between-community diversity.}
With two planted communities, the variance of stance across the
population decomposes exactly into a \emph{within}-community component
(the size-weighted mean of the two block variances) and a
\emph{between}-community component (the size-weighted variance of the
two block means). Figure~\ref{fig:within_between} tracks both through
the campaign and quiet rounds in the segregated cell
($\lambda_{\mathrm{r}}{=}4$, $\alpha{=}1$) on three topics --- the
reference topic, vaccines, and MMR, on which both campaigns argue
against the claim (Appendix~\ref{app:validity}) --- each run on the
same sixteen graphs with four conditions at $n{=}16$. In the
uninfluenced population the between component falls by about a third
to a half over the run (decoupling $0.20\!\to\!0.13$, vaccines
$0.18\!\to\!0.09$, MMR $0.23\!\to\!0.12$) while the within component
falls less, so the two communities drift toward each other even
without a campaign. Campaigns act almost entirely on the between
component: relative to the seed-matched null at round~33 they remove a
further $0.05$--$0.08$ of between-community variance in every
condition on every topic except the \chal{} alone on decoupling
($-0.03$), while changing the within component by at most $0.03$ and
usually by less than $0.01$. In the segregated population, campaigning
of either kind therefore closes the gap between the two communities
and leaves the diversity inside each community where it was. The
control cells confirm that the between component measures structure:
with unsorted opinions ($\alpha{=}0.5$) or no communities
($\lambda_{\mathrm{r}}{=}1$) it is $0.02$ or less throughout.

\clearpage

\subsection{Baseline drift and baseline diversity}
\label{app:baseline}

\begin{figure*}[tp]\centering
\appfig{0.8\textwidth}{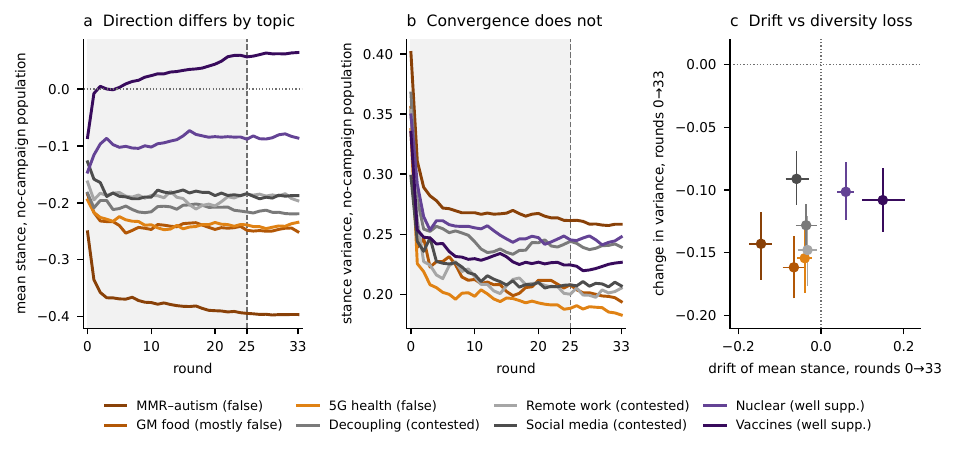}
\caption{The uninfluenced population on each of the eight topics
($n{=}20$ runs per topic, means over runs). (a)~Mean stance by round.
(b)~Stance variance by round. (c)~Change in mean stance against change
in variance from round~0 to round~33, one point per topic with 95\%
bootstrap CIs over runs. Campaign rounds are shaded although no campaign
is active; the dashed line marks where the quiet rounds begin in the
campaign conditions.}
\label{fig:baseline}
\end{figure*}

Two baseline phenomena coexist in the no-campaign population and are
separable. \emph{Directional drift} --- the change in mean stance ---
differs by topic in sign and size, from $-0.15$ on MMR to $+0.15$ on
vaccines, and is distinguishable from zero on all eight topics. Loss of
\emph{diversity} --- the change in stance variance --- does not: every
topic's uninfluenced population ends with $55$--$71$\% of its
starting variance (change $-0.16$ to $-0.09$), and the size of the loss
is unrelated to the size or sign of the drift (correlation of $|$drift$|$
with variance change across topics $0.17$). In this simulated LLM
population, where the baseline goes depends on the claim; that it
converges does not. These are population-level regularities of the
simulated society and are not offered as claims about human
populations.

\subsection{Polarisation}
\label{app:polarisation}

\begin{figure*}[tp]\centering
\appfig{0.8\textwidth}{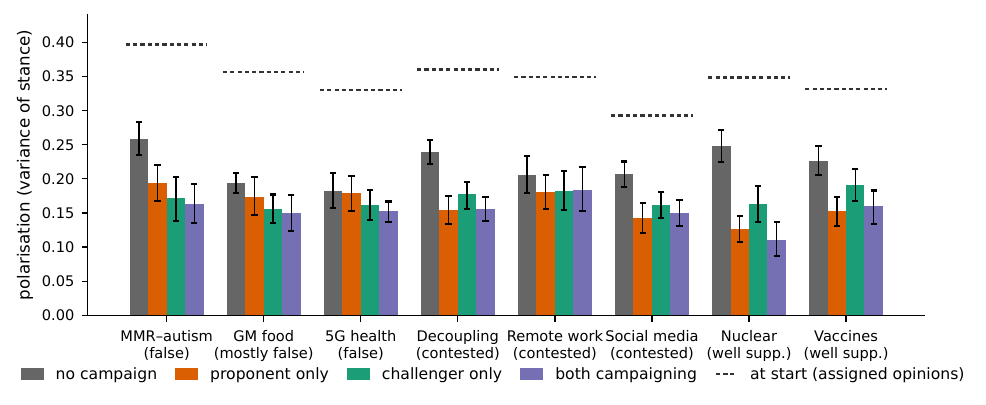}
\caption{Final polarisation (variance of stance at round~33) under each
condition, per topic, with 95\% bootstrap CIs. The dashed line is the
polarisation of the assigned starting attitudes.}
\label{fig:polarisation}
\end{figure*}

The population starts deliberately spread and converges even when left
alone. Every campaign condition ends less polarised than the
uninfluenced population, in all eight topics and in all four network
cells: a campaign gives everyone it reaches the same argument, and the
agents it moves most are those furthest from it, which compresses the
distribution regardless of the direction in which the mean moves. We
therefore find no evidence, at this scale and from this starting
distribution, that adversarial campaigns or community structure increase
polarisation.

\begin{figure*}[tp]\centering
\appfig{0.8\textwidth}{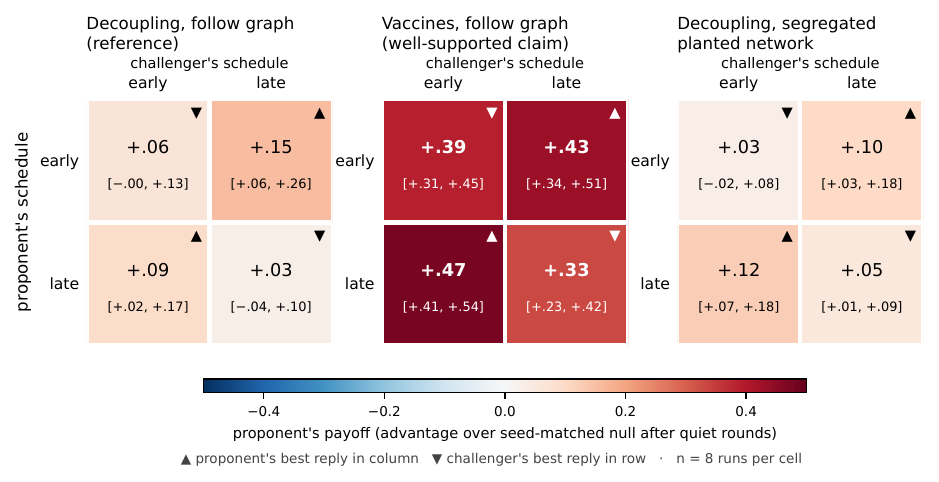}
\caption{The four corner cells of the timing matrix (early and late
schedules for each side; $n{=}8$ per cell) on the reference topic
(left), on vaccines (centre) and on the reference topic with the
segregated planted network of Appendix~\ref{app:planted} (right).
Cells report the \prop{}'s payoff after the quiet rounds with 95\%
bootstrap CI; $\blacktriangle$ and $\blacktriangledown$ mark best
replies as in Figure~\ref{fig:payoff}.}
\label{fig:timing_robust}
\end{figure*}

\begin{figure*}[b]\centering
\appfig{0.7\textwidth}{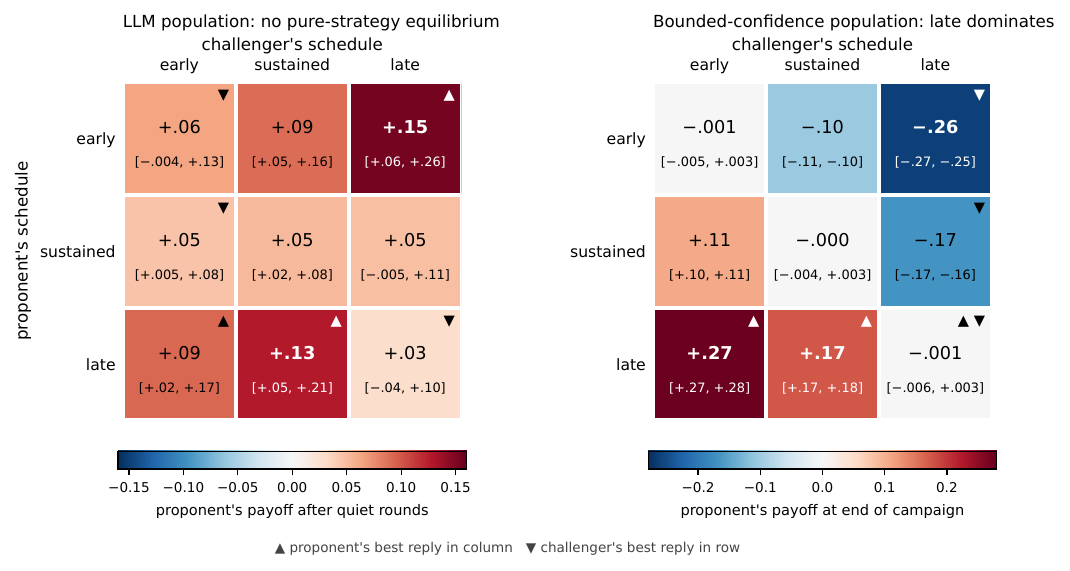}
\caption{The two matrices of Figure~\ref{fig:payoff} with per-cell 95\%
bootstrap CIs ($n{=}8$; $n{=}16$ for sustained--sustained). The
numerical panel is shown at the end of the campaign because retention
toward the starting attitude ($\lambda{=}0.3$ per round) erases almost
all of its campaign effect within the eight quiet rounds (largest
after-quiet payoff $0.016$); the best-reply structure is identical at
both measurement points.}
\label{fig:bcm}
\end{figure*}

\subsection{Robustness of the timing result}
\label{app:timing_robust}

In all three settings the \prop{}'s best reply avoids the \chal{}'s
schedule and the \chal{}'s coincides with it, so no cell is a best reply
for both. The paired contrast between off-diagonal and on-diagonal
payoffs is $+0.076$ $[+0.009,+0.142]$ on the reference, $+0.092$
$[+0.032,+0.153]$ on vaccines and $+0.071$ $[+0.002,+0.137]$ on the
segregated network (bootstrap over the eight seeds that carry all four
cells). The levels differ --- on vaccines every cell exceeds $+0.33$
because both campaigns raise acceptance of a well-supported claim ---
but the ordering within the matrix is the same. The checks are limited
to the corner cells and to one additional topic and one additional
network; the sustained schedule and the remaining six topics were not
re-run.

\subsection{Numerical comparison arm}
\label{app:bcm}
Agents hold one number $x_i\in[-1,1]$ and update by
$\tilde{x}_i^{t}=x_i^{t}+\mu\big(|\mathcal{C}_i^{t}|^{-1}\sum_{j\in\mathcal{C}_i^{t}}x_j^{t}-x_i^{t}\big)$
over the neighbours within $\varepsilon$ of them, then
$x_i^{t+1}=(1-\lambda)\big(\tilde{x}_i^{t}+\eta\sum_{c:\,i\in\mathcal{T}_c^{t}}\sigma_c\big)+\lambda x_i^{0}$,
with $\sigma_{\mathrm P}{=}{+}1$, $\sigma_{\mathrm C}{=}{-}1$,
$\varepsilon{=}\mu{=}\lambda{=}0.3$ and $\eta{=}0.1/(1-\lambda)$ so one
delivery moves a neutral agent by $0.1$. Each delivery logged in the LLM
runs is replayed to the same agent in the same round; no API call is
made. The numerical game has a unique pure-strategy equilibrium at
late--late (present in $100\%$ of bootstrap resamples at the end of the
campaign and $94\%$ after the quiet rounds), the matrix is
antisymmetric, and the equilibrium is unchanged for
$\lambda\in\{0,0.1,0.3\}$ and $\varepsilon\in\{0.3,0.6,1.0\}$: with
symmetric, memoryless pushes, whoever speaks last wins. The LLM
population has no pure-strategy equilibrium ($81\%$ of resamples) and
an all-positive matrix. Timing cycles because of how LLM agents revise,
not because of the allocation game itself.